\documentclass[twoside,fleqn]{article}
\usepackage{espcrc2,epsfig,graphicx}
\newcommand{\E}[1]{$10^{#1}$eV}

\newcommand{\text}[1]{\mbox{#1}}

\title{Comment on ``The Lamb Shift and Ultra High Energy Cosmic Rays'' and Comment on ``Vacuum Polarization Energy Losses of High Energy Cosmic Rays''.}
\author{Olivier Deligny
\address{L.P.N.H.E. Paris VI-VII, 4 place Jussieu, 75252 Paris, France} 
}
\begin{document}
\begin{abstract}
The cosmic ray spectrum has been shown to extend well beyond \E{20}. With nearly 20 events observed
in the last 40 years, it is now established that particles are accelerated or produced in the
universe with energies near or above \E{21}. No nearby astrophysical object has been shown to correlate with 
the arrival directions of the highest energy events, yet the exponential cut-off in the high energy end of the 
spectrum one expects to see in the case of far sources is not visible.
It was recently pointed out that the influence of the vacuum of quantum electrodynamics on particle propagation 
could explain qualitatively this mystery. This note is a critic to these ideas.
\vspace{1pc}
\end{abstract} 
\maketitle

\section{Introduction}
The origin of Ultra High Energy Cosmic Rays observed on Earth is a long lasting 
mystery\cite{Yoshida,Sigl,Bertou,Nagano}.
While the cosmic ray spectrum is now shown\cite{HiresTaup99,Agasa00} to extend beyond 
\E{20}, mechanisms producing or accelerating particles with energies near or above \E{21} are still uncertain. 
The detection of a large flux of events above $5\times$\E{19} is considered with great interest because of the absence of 
the Greisen-Zatsepin-Kuzmin (GZK) cut-off\cite{GZK}. This cut-off should be observed if cosmic primaries are 
protons originating from cosmologically distributed sources. 
Above $5\times$\E{19}, on their way from the sources to Earth, cosmic ray protons loose their energy photo-producing 
pions against the cosmic microwave background (CMB). Therefore sources further than about 50 Mpc away from Earth are not expected to contribute to the high energy end of the spectrum giving rise to the GZK cut-off. 
If sources are closer than 50 Mpc, protons of \E{20} should point toward them, however the present data does not 
indicate any correlation between arrival direction of these events and local distribution of galaxies and is 
mostly isotropic. Many ideas have been put forward in the literature to answer some of those problems, 
but the general opinion is that none of them are complety satisfactory. Recently, two papers\cite{Xue,Maisheev} 
analysing the propagation of protons in vacuum and in the CMB claimed to qualitatively explain the absence 
of the GZK cut-off. After recalling a famous analogy between the dielectric properties of matter and 
the quantum electrodynamics vacuum, we discuss these ideas and demonstrate that one leads to 
(unexplicit!) Lorentz violation while the result of the semi-classical calculation of the other 
is not compatible with a proper QED treatment.  

\section{Dielectric properties of matter}

When an external electric field is applied to a medium, the atomic electrons are shifted from their 
original equilibrium position in the field of the nuclei. 
Similarily if we introduce a negatively charged particle in this medium, 
atomic electrons will be pulled away and an excess of positive charges will surround the particle.
Thus, the negative charge is screened by the positive ones and the force felt by two charges 
introducted in the medium becomes
\[
\frac {e^2}{\epsilon r^2}
\]
where $r$ is the distance between the two charges and $\epsilon$ is the dielectric constant ($\epsilon > 1$).
This description applies to distances greater than the interatomic distances. 
For smaller distances there is no screening. A unified definition was introduced by Landau\cite{Landau} 
through the integral operator $\epsilon (r)$, which approachs 1+ for $r$ small
compared to the interatomic distance and the classical $\epsilon$ at larger ones.

\section{Vacuum polarization}
The classical relativistic vacuum is Lorentz invariant. It is a reference system for the description of the motion as well as the propagation of the electromagnetic field. This vacuum is empty.
This classical description was modified with the birth of statistical mechanics, and of course of quantum physics. 
From the statistical mechanics point of view, at non-zero temperature, the space is filled by 
blackbody radiation. In this approach, extracting all matter in a region of space is not sufficient 
to turn it into vacuum~: one must also bring the temperature down to zero. 
The quantum theory goes further and predicts that field fluctuations subsist even at zero temperature. 
This is a consequence of the definition of the quantum vacuum and of the Heisenberg inequalities.
Vacuum is the state for which there is no field excitations but which still contains zero-point field fluctuations. 
These fluctuations have observable consequences as the spontaneous emission of an isolated atom, the Lamb shift, or 
the Casimir force... 
\par
It is instructive to consider the vacuum of quantum electrodynamics as a polarized medium under the effect of an external field, in order to understand how this vacuum feels the application of this external field. 
This well known analogy is quite rigourous for the electric field even from the point of view of QED. 
We can think of this vacuum as a stock of virtual particles which can have an existence 
during short periods of time according to the Heisenberg inequalities. Thus, when, for instance, 
a positive charge is introducted in this stock, polarization of the virtual pairs occurs~: the virtual positrons 
are expelled from the positive charge whereas the virtual electrons are attracted by the charge. A positive vacuum 
surrounds the charge, and its bare value depends on the way it is measured, that is to say depends on the radius of the sphere we take around the charge to measure the surface value of the electric field. 
Again, screening occurs up to a distance $r$ of the order of $\hbar /mc$ where $m$ is the mass of the (screening) 
particle pairs.

\section{Can the vacuum accelerate a charged particle ?}

In his paper relating the Lamb shift and the propagation of protons in vacuum\cite{Xue}, the author considers 
that the proton field must be coupled to the vacuum fluctuations. In order to describe the propagation of the proton in the QED vacuum, fundamental fields are decomposed following their classical part and their quantum fluctuations parts; leading to the Lagrange density $L(x)$~:
\begin{eqnarray*}
L(x)&=& -\frac{1}{4}F^2 - \frac{1}{4}F_q^2 + \overline{\Psi}(i\gamma^{\mu}\partial_{\mu}-m_p+e\gamma^{\mu}A_{\mu}^q)\Psi\\
&+&\overline{\Psi}_q(i\gamma^{\mu}\partial_{\mu}-m_e+e\gamma^{\mu}(A_{\mu}+A_{\mu}^q))\Psi_q + (c.t.)
\end{eqnarray*}
where the fields indexed by $q$ describe the quantum fluctuations and (c.t.) are all necessary counterterms. At this level, we wonder why the terms $-\frac{1}{2}FF_q$ and $e\overline{\Psi}\gamma^{\mu}A_{\mu}\Psi$ are absent of the decomposition~: there is of course no reason to drop these terms. 
Next, an effective Lagrange density is computed by perturbation in term of the electromagnetic coupling:
\begin{eqnarray*}
L_{eff}(x)&=&-\frac{1}{4}F^2 + \overline{\Psi}(i\gamma^{\mu}\partial_{\mu}-m_p)\Psi\\
&+& \emph{tr}\ln [S_F^{-1}(x)-V(x)]
\end{eqnarray*}
$V(x)$ is arbitrary choosen to be the same than the potential which describes the propagation of a lepton in a classical external field\cite{Itzykson}, whereas the needed diagrams to get this potential are not permitted by the starting Lagrange density because of the absence of the term $e\overline{\Psi}\gamma^{\mu}A_{\mu}\Psi$.
From there, the main argument is based on the fact that the difference of energy of the vacuum spectrum in presence of the proton can be transferred to the proton. This difference occurs because the virtual pairs are polarized along the proton path whereas these fluctuations are totally random in the absence of the electromagnetic field of the proton. According to the author, this difference \emph{accelerates} the proton. In order to estimate this acceleration, he considers that virtual fermion pairs can be treated as bound states of size $\frac {1}{\alpha m}$ and density $\frac{3}{4\pi}\alpha ^3 m^3$. At this point, the explicit form of the Hamiltonian of such bound states (not well funded!) is not developped at all, and the compact pile of these bound states to estimate their density is rather surprising. Introducing a modified relation of dispersion to take into account the energy gain responsible of the acceleration 
\begin{eqnarray*}
E + \delta E = \sqrt{(\vec{p}+\delta \vec{p})^2 + m_p^2}
\end{eqnarray*}
the author finally obtains the numerical estimate of the energy gain (without loss of generality, we can consider a straight-line motion)
\begin{eqnarray*}
\frac{\delta E}{\delta x} = \frac{\delta p}{\delta \tau} \sim 2.25\times 10^{-5} \frac{v}{c}\, eV cm^{-1}
\end{eqnarray*}
However, quantum field theory is based on the theory of the relativity. The properties of the quantum vacuum must satisfy to a precise relation with respect to the principles of Einstein's theory. Thus, in order to satisfy the principles of relativity, it is essential that the vacuum possesses the properties required by the relativity, that is to say the vacuum must not distinguish inertial frames. If the force that the vacuum exerts on a particle moving with a uniform velocity is not zero, then the vacuum distinguishes an inertial frame with respect to a rest frame. Of course, quantum electrodynamics predict that this force vanishes. In general, this theory predicts that vacuum fluctuations seen by two inertial observers are the same, simply because the vacuum is Lorentz invariant. The acceleration by the vacuum described in\cite{Xue} depends on the speed of the particle and so, doesn't preserve the properties of the vacuum for any inertial observer.The model, which makes no sense from the starting Lagrange density, \emph{is not Lorentz invariant}.
\par
To go a little further about this subject, let us consider now general motion in vacuum. Fulling and Davies have shown\cite{Fulling} that the vacuum resists to any motion to take it back to a \emph{uniformly accelerated} motion. The corresponding energy variation is radiated. Other remarkable works of Jaekel and Reynaud\cite{Jaekel} point out that for a uniformly accelerated motion, characterised by the acceleration
\begin{eqnarray*}
w^{\mu} = \ddot{v}^{\mu} + v^{\mu}\dot{v}^{\mu 2}
\end{eqnarray*}
where 
\begin{eqnarray*}
v^{\mu}(\tau) = \frac{dx^{\mu}}{d\tau}
\end{eqnarray*}
and $\tau$ is the proper time satisfying $d\tau ^2 = dx ^2$, the so-called Abraham-Lorentz reaction force, which is proportional to $w^{\mu}$, vanishes. Properties of quantum vacuum are therefore invariant for a set of transformations which include uniformly accelerated frames. These are the conformal transformations.\\

\section{A forgotten damping force to explain the cosmic rays spectrum?} 
\par
In another paper describing the energy losses with a damping force by vacuum
polarization\cite{Maisheev}, a second author considers the propagation of protons in
the cosmic microwave background and thus evades our last critic about the impossibility of acceleration
or deceleration in vacuum because of the introduction of a preferred frame. 
The author calculates the energy losses by
vacuum polarization in the same way used by Landau for the ionisation losses of a
charged particle travelling in a continuous medium\cite{Landau}.
Even if the damping force considered is caused by electric field of virtual pairs, his approach is classical, consisting in the introduction of a permittivity tensor of the gas of photons (the author assumes that the connection between polarization and permittivity was established by Baier and Katkov\cite{Baier}). 
This approach allows him to consider QED vacuum and matter
in a unified point of view. After the resolution of the Maxwell equations mapped
on what is usually done in a continuous medium, the author computes the final
damping force~:
\begin{eqnarray*}
F \sim e^6 f(v^2)
\end{eqnarray*}
where $f$ is a function which contains the proton form factor.
The numerical estimation shows that this force becomes essential in the
energy range between the knee of high energy cosmic rays spectrum and the GZK
cut-off [\E{15}-\E{20}]. Taking into account this energy loss, these new numerical results allow the author to a new interpretation of the existence of the knee, a new explanation of the absence of a clear photoabsorption threshold in the spectrum of high energy cosmic rays, as well as the observed changes in the composition of primaries in this energy range (when applied to iron nuclei).
\par However, if this approach is made in the framework of the quantum electrodynamics as
required (only the electric field of virtual pairs can cause this damping
force), one observes that the computation of the damping force leads to
an expression of order 3, which is compatible with the interference term between lower order
of inverse Compton scattering and correction with
vacuum polarization and internal photon line in the perturbation theory. In the incoherent gas of photons, this process should lead to a correction to the lower order of the inverse Compton scattering between protons (or nuclei) and photons. This Compton interaction is known to lead to only a negligibly small energy loss in this energy range\cite{Stecker}. Therefore, this semi-classical approximation doesn't seem applicable in this context.
 
\section{Conclusions}

Two recent ideas, proposed to explain the mysterious absence of the GZK cut-off
in the UHECR spectrum, have been discussed. We have shown that the first cannot
account for the observed phenomenon as the first approach does not respect
fundamental principles of relativity. The second is based on a semi-classical approximation
which describes coherent effects and doesn't seem to be applicable in the context of propagation
of UHECR.
Consequently,
the ultra high energy cosmic rays fluxes remains a mystery that current
generation of detectors will hopefully solve in the near future\cite{Auger}. 

\section*{Acknowledgments}

Many thanks to P.Billoir, M.Boratav, J.Hirn, A.Letessier-Selvon and E.Parizot for helpful discussions, comments and careful read of the text.

\font\nineit=cmti9
\font\ninebf=cmbx9

\end{document}